\documentclass[%
 reprint,
superscriptaddress,
 amsmath,amssymb,
 aps,
]{revtex4-2}

\usepackage{graphicx}
\usepackage{dcolumn}
\usepackage{bm}

\begin{document}

\preprint{}

\title{Generation of intense cylindrical vector beams by Faraday effect in plasma}

\author{Wei Liu}
\affiliation{%
Department of Plasma Physics and Fusion Engineering, University of Science and Technology of China, Hefei, Anhui 230026, People's Republic of China
}
\author{Qing Jia}
\email{qjia@ustc.edu.cn}
\affiliation{%
Department of Plasma Physics and Fusion Engineering, University of Science and Technology of China, Hefei, Anhui 230026, People's Republic of China
}

\author{Jian Zheng}
\affiliation{%
Department of Plasma Physics and Fusion Engineering, University of Science and Technology of China, Hefei, Anhui 230026, People's Republic of China
}
\affiliation{Collaborative Innovation Center of IFSA, Shanghai Jiao Tong University, Shanghai 200240, People's Republic of China
}


\date{\today}

\begin{abstract}
Cylindrical vector (CV) beams, whose polarizations are cylindrically symmetric, have recently been widely applied in high energy density physics such as electron acceleration and intense spatiotemporal optical vortices generation. Thermal-damage-resistant plasma optics are expected to generate intense CV beams. In this work, based on the Faraday effect, we propose a method that can directly convert an intense linearly/circularly polarized Gaussian beam into a CV/vortex beam by setting up an azimuthally distributed axial magnetic field in the plasma. Three-dimensional particle-in-cell simulations demonstrate good conversion efficiency, which offers a new degree of freedom for manipulating high-power laser pulses and paves the way for further studies on ultra-strong vector beams. In addition, our work reveals a new possible source of photon orbital angular momentum related to magnetized plasma in astrophysics and space physics. 
\end{abstract}

\maketitle



Polarization is one of the fundamental properties of electromagnetic wave. In laser plasma physics, there are many processes that are closely related to laser polarization, such as laser resonance absorption\cite{Freidberg1972,Woo1978}, axial self-generated magnetic field\cite{Deschamps1970,Haines2001}, high harmonic generation\cite{Corkum1993,Bulanov1994,Lichters1996,Quere2006} and electron acceleration\cite{Liu2013}. In these works, electromagnetic waves are typically considered linearly or circularly polarized, whose polarizations are uniformly distributed in the transverse plane. Exotic laser beams with transversely non-uniformly distributed polarization, namely, vector beams\cite{Rubinsztein2016}, have attracted much attention in the past two decades. Cylindrical vector (CV) beams\cite{Zhan2009} are a special class of vector beams whose polarizations are cylindrically distributed. Representative radially polarized (RP) and azimuthally polarized (AP) electromagnetic waves have been studied extensively mainly due to their unique focusing characteristics\cite{Quabis2000,Dorn2003}. It has been proven that RP beam can be focused more tightly than linearly polarized beam. Near the focal plane, the tightly focused RP(AP) beam will result in strong longitudinal electric (magnetic) fields near the optical axis.

Weak CV beams have been employed in fields such as super-resolution imaging\cite{Kozawa2018} and optical trapping\cite{zhan2004}. Recently, intense CV beams have been of interest in laser plasma physics and have brought some new effects. Due to the strong longitudinal electric fields, RP beam is applied in the direct laser acceleration of electrons\cite{Zaim2017,zaim2020}. In particular, when using a plasma mirror as an electron injector, electron bunches with MeV energies and hundreds of pC charges are obtained\cite{zaim2020}. In addition, it demonstrates that intense spatiotemporal optical vortices\cite{Bliokh2012,Chong2020} can be generated by intense RP beams obliquely reflected from a solid-plasma surface\cite{Chen2022}.

However, as far as we know, the generation of intense CV beams remains challenging. Conventional methods for generating intense CV beams\cite{Pohl1972,Machavariani2008,Chen2014} might be restricted due to the optical thermal threshold. Plasma, benefiting from sustaining very high light intensity, has recently been widely used to manipulate intense lights, such as plasma mirrors to improve the temporal contrast of femtosecond pulses\cite{Thaury2007}, plasma-based options to generate vortex beam\cite{Shi2014,Vieira2016,Leblanc2017,QuKenan2017,Long2020}, holographic plasma lenses to focus intense lasers\cite{Edwards2022}, plasma photonic crystals\cite{lehmann2016} and so on. In this study, utilizing the Faraday effect\cite{Faraday1846}, we propose a straightforward and effective plasma-based approach for converting an intense linearly/circularly polarized Gaussian beam into a CV/vortex beam.

The investigation of plasma-based generation of vector beam and vortex beam in the laboratory also holds significant importance for observations in astrophysics and space physics. Currently, our understanding of the universe is primarily derived from observations of electromagnetic waves. It is crucial to extract as much information as possible from the collected electromagnetic waves, given the scarcity of data and the high cost. Recently, the photon orbital angular momentum (POAM)\cite{Allen1992}, which is linked to the vortex beam, is increasingly attracting attention as a new degree of freedom in the field of astronomy\cite{Harwit2003,Berkhout2008,Tamburini2011,Tamburini2019,Tamburini2021}. In 2003, Harwit outlined several potential sources of astrophysical POAM\cite{Harwit2003} from radiation emitted by luminous pulsars and quasars to the cosmic microwave background radiation. Especially, a recent study\cite{Tamburini2019} has confirmed the presence of POAM generated by the gravitational effect near rotating black holes, as previously proposed by \citet{Tamburini2011}. And this research demonstrates that POAM can be directly employed to measure the rotation parameters of a black hole. It is worth noting that plasmas found in the universe and celestial bodies are often magnetized, such as the plasmas in magnetic reconnection process\cite{Yamada2010}. We are curious whether POAM can be generated after electromagnetic waves passing through these magnetized plasmas and whether they can be used for relevant observations.


The difference in dispersion between left- and right-hand circularly polarized electromagnetic waves in axially magnetized plasma is a well-established phenomenon that gives rise to distinct phase and group velocities\cite{Chen2012}. Taking into account the difference in \textit{group velocity}, simulations have shown that a linearly polarized laser pulse passing through an axially magnetized plasma can split into two laser pulses with opposite circular polarizations, provided the axial magnetic field is strong enough\cite{Weng2017}. Considering the differences in \textit{phase velocities}, the polarization of a linearly polarized laser undergoes rotation after passing through an axially magnetized plasma, which is referred to as the Faraday effect\cite{Chen2012}. We will now provide a brief overview of the Faraday effect and offer a brief introduction to the generation of CV beam based on it.

The propagation of left- and right-hand circularly polarized electromagnetic waves along a weakly magnetic field ($|{{{\omega }_{ce}}}/{\omega }|\ll 1$) in a plasma is associated with different dispersions, which can be expressed\cite{Chen2012}
\begin{equation}
\label{eq:num1}
{N}_{L/R}=\frac{kc}{\omega }\approx \sqrt{1-\frac{{{n}_{e}}}{{{n}_{c}}}}\pm \frac{1}{2}\frac{1}{\sqrt{1-\frac{{{n}_{e}}}{{{n}_{c}}}}}\frac{{{n}_{e}}}{{{n}_{c}}}\frac{{{\omega }_{ce}}}{\omega},
\end{equation}
where $N_{L/R}$ is the refractive index for left/right-hand circularly polarized electromagnetic wave, $k$ is the laser wavenumber in magnetized plasma, $c$ is the speed of light in vacuum, $\omega$ is the laser frequency, ${{n}_{e}}$ is the electron number density, ${{n}_{c}}={{{\varepsilon }_{0}}{{m}_{e}}{{\omega }^{2}}}/{{{e}^{2}}}$ is the critical number density, ${{m}_{e}}$ is the electron mass, ${{\omega }_{ce}}={e{{B}_{x}}}/{{{m}_{e}}}$ is the electron cyclotron frequency,  $e$ is the elementary charge, and ${{B}_{x}}$ is the axial magnetic field.

The incident linearly polarized electromagnetic wave can be represented as ${\bm{E}}_{in}=[({{\bm{e}}_{\bm{y}}}-i{{\bm{e}}_{\bm{z}}})+({{\bm{e}}_{\bm{y}}}+i{{\bm{e}}_{\bm{z}}})]{{E}_{0}}\exp[i({{k}_{0}}x-\omega t ) ]$, where ${{\sigma }_{x}}=\pm 1$ in ${{\bm{e}}_{\bm{y}}}+{{\sigma}_{x}}i{{\bm{e}}_{\bm{z}}}$ represent right/left-hand circularly polarized and ${{k}_{0}}$ is the laser wavenumber in vacuum. After propagating distance $L$, the output wave can be written as
\begin{equation}
{{\bm{E}}_{out}}=2{{E}_{0}}({\cos\varphi}{{\bm{e}}_{\bm{y}}} + {\sin\varphi} {{\bm{e}}_{\bm{z}}})\exp ( i \Phi -i\omega t ),
\label{eq:num2}
\end{equation}
where $\varphi={k}_{0}L({N}_{L}-{N}_{R})/{2}$, $\Phi={k}_{0}L({N}_{L}+{N}_{R})/{2}={{k}_{0}}L\sqrt{1-{{{n}_{e}}}/{{{n}_{c}}}}$. The output wave is still linearly polarized but with polarization rotation angle
\begin{equation}
\label{eq:num3}
\varphi \approx \frac{1}{2}\frac{1}{\sqrt{1-\frac{{{n}_{e}}}{{{n}_{c}}}}}\frac{{{n}_{e}}}{{{n}_{c}}}\frac{{{\omega }_{ce}}}{\omega }{{k}_{0}}L.
\end{equation}
The Faraday rotation angle is directly linked to plasma density, plasma length, and the strength of the axial magnetic field. In most cases, researchers focus on studying a plasma with uniform magnetization in the transverse plane, which ensures that the output beam remains linearly polarized. However, this uniformity restricts the degree of polarization manipulation that can be achieved. In cases where the magnetization in the transverse plane is non-uniform, the incident linearly polarized beam undergoes a transformation into a vector beam.  By skillfully designing the distribution of the magnetic field, it becomes possible to generate a vector beam with desired characteristics, such as a CV beam.

\begin{figure}
\centering\includegraphics[]{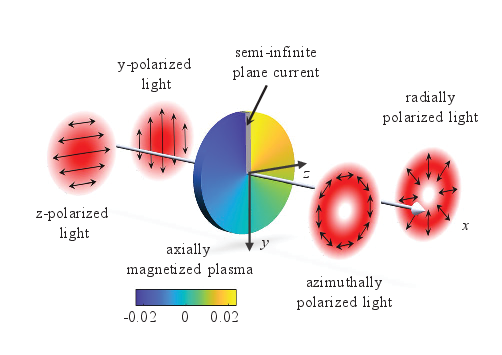}
\caption{\label{fig:num1}Illustration of a linearly y/z-polarized beam transformed into a RP/AP beam after passing through an axially ($x$) magnetized plasma (normalized by ${{m}_{e}}\omega /e$). }
\end{figure}

The simplified distribution of a RP beam at the waist can be written as\cite{Zhan2009}
\begin{eqnarray}
{\bm{E}}={{E}_{0}}{\left( \frac{r}{{{w}_{0}}} \right)\exp \left( -\frac{{{r}^{2}}}{{{w}_{0}}} \right)} \exp[i({{k}_{0}}x-\omega t )]{{\bm{e}}_{\bm{r}}},
\label{eq:num31}
\end{eqnarray}
where ${w}_{0}$ is the beam waist, $r=\sqrt{{{y}^{2}}+{{z}^{2}}}$ and $\theta =\arctan \left( {z}/{y}\right)$. A significant difference between a RP beam and a linearly polarized beam is that the polarization of a RP beam varies across different spatial locations, but all along the radial direction. In order to convert linearly polarized beam into RP beam, it is necessary to introduce spatially varying Faraday rotation angles $\varphi(\theta)$. We have discovered that by setting the Faraday rotation angle in Eq.~(\ref{eq:num3}) as $\varphi \left( r,\theta  \right)=\theta$, the incident laser beam with linear polarization in the $y$ direction can be converted into RP beam, as illustrated in Fig.~\ref{fig:num1}. By altering the polarization direction of the incident beam, other CV beams like AP beam can be obtained in principle.

Furthermore, in accordance with Eq.~(\ref{eq:num2}), the Faraday rotation not only induces the rotation of laser polarization but also introduces an additional phase $\Phi$. In order to prevent the introduction of a spiral phase $\exp(i\theta)$ for CV beams\cite{Zhan2009}, the phase $\Phi$ should be independent of azimuth, which implies that $\varphi \left( r,\theta  \right)=\theta$ can be achieved by setting an azimuthally distributed axial magnetic field ${{B}_{x}}={{B}_{ext}}\theta$ with 
\begin{equation}
\label{eq:num4}
{{B}_{ext}}=\frac{2{{m}_{e}}\omega {n}_{c} \sqrt{1-\frac{{{n}_{e}}}{{{n}_{c}}}}}{ e{n}_{e}{{k}_{0}}L}.
\end{equation}
This magnetic field distribution, depicted in Fig.~\ref{fig:num1}, can be  generated approximately by a semi-infinite plane current. Further details and a comprehensive discussion on this topic will be provided in Sec. \ref{sec:chap4}.


\begin{figure}[b]
\centering\includegraphics[]{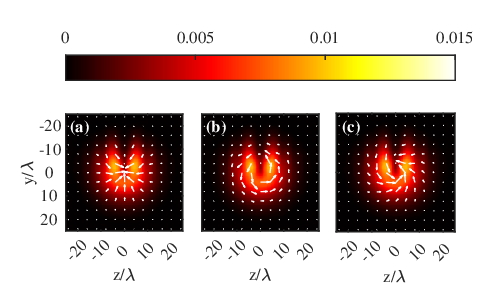}
\caption{\label{fig:num2}PIC simulation results of laser amplitude (normalized by ${{m}_{e}}\omega c/e$) and electric field vector distributions after different linearly polarized beams pass through the ${{B}_{x}}={{B}_{ext}}\theta $ magnetized plasma at $x=65\;\rm{\mu m}$. (a) RP beam generated by a y-polarized Gaussian beam; (b) AP beam generated by a z-polarized Gaussian beam; (c) A generalized CV beam generated by linear superposition of y-polarized and z-polarized Gaussian beams. The background color scale shows the laser amplitude, and the white arrows represent the electric field vectors. }
\end{figure}

To verify the above scheme, three-dimensional (3D) particle-in-cell (PIC) simulations are conducted with the code Smilei\cite{Derouillat2018}. A linearly polarized Gaussian beam with wavelength $\lambda =1\rm{\mu m}$ (in vacuum) and radius of waist ${{w}_{0}}=10\rm{\mu m}$ normally incidents on a plasma located at $15\rm{\mu m}<x<65\rm{\mu m}$. The laser intensity rises linearly over 10 laser periods to its maximum intensity ${{I}_{0}}=1.37\times {{10}^{14}}\rm{{W}/{c{{m}^{2}}}}$ (corresponding to $a_0=eE_0/m_e\omega c =0.01$) and then remains constant. The simulation box is $80{\rm{\mu m}}(x)\times 50{\rm{\mu m}}(y)\times 50{\rm{\mu m}}(z)$ with $1280\times 400\times 400$ cells. The cell sizes are $\mathrm{d}x ={\lambda }/{16}$, $ \mathrm{d}y=\mathrm{d}z={\lambda}/{8}$. Each cell has applied 16 particles for electrons, and the ions are set to be immobile. The electrons are uniformly distributed with number density ${{n}_{e}}=6.27\times {{10}^{26}}\,\rm{{m}^{-3}}$, corresponding to $0.56n_c$. The axial magnetic field distribution is set as Eq.~(\ref{eq:num4}) with ${{B}_{ext}}$=$80.17\,{\rm{T}}$, which satisfies $\left| {{{\omega }_{ce}}}/{\omega } \right|\ll 1$. It should be noted that in the simulations, such magnetic field only participates in pushing the particles but does not participate in the Maxwell solver\cite{Derouillat2018}, which can be regarded as the steady magnetic field generated by an external current or kinds of magnets and is physically reasonable.

Figure~\ref{fig:num2} presents the laser amplitude and electric field vector distributions when lasers exiting the magnetized plasma for different linearly polarized incident Gaussian beams. Figure~\ref{fig:num2}(a) displays the result of a y-polarized Gaussian beam passing through the plasma. The laser intensity is modulated to have a null intensity in the center, which is a key feature of the CV beam. Besides, judging from the distribution of the laser electric field vectors represented by the small white arrows, the linearly polarized beam is successfully converted into a RP beam. Similarly, the result of the transformation of a z-polarized Gaussian beam into an AP beam is shown in Fig.~\ref{fig:num2}(b). In general, an arbitrary CV beam can be generated by a reasonable linear superposition of y-polarized and z-polarized Gaussian beams, as indicated in Fig.~\ref{fig:num2}(c). 

Note that in Fig.~\ref{fig:num2}, the intensity of the laser is significantly modulated near the line of $z=0$ ($y<0$). In our theoretical  scheme, the axial magnetized plasma is regarded as an optical medium with the dispersion relation shown in Eq.~(\ref{eq:num1}), which requires the scale length of the electron motion (${{L}_{ele}}\propto{{a}_{0}}\lambda$ when ${{a}_{0}}<1$) to be much smaller than the scale length of the magnetic field (${{L}_{mag}}\sim| {r{{B}_{x}}}/{({\partial {{B}_{x}}}/{\partial \theta })}|$). This condition is satisfied in most areas except the region near the line of $z=0$ ($y<0$), where a large gradient in the magnetic field exists as revealed in Fig.~\ref{fig:num1}. This large magnetic gradient leads to non-typical motions of electrons, which significantly modulates the laser intensity as shown in the PIC simulations. However, these mode imperfections will gradually decrease as the laser exits the plasma and propagates in vacuum.

\begin{figure}[t]
\centering\includegraphics[]{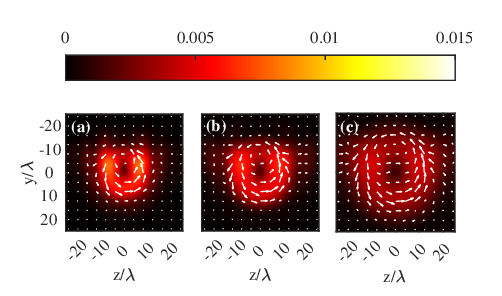}
\caption{\label{fig:num3}Distributions of electric field vectors and laser amplitude (normalized by ${{m}_{e}}\omega c/e$) as the RP beam leaves the plasma and propagates at different distances in vacuum: (a) $\Delta x=50\rm{\mu m}$, (b) $\Delta x=100\rm{\mu m}$, (c) $\Delta x=200\rm{\mu m}$. The distance $\Delta x$ is referenced by $x=65\rm{\mu m}$, where the laser exits the plasma. These results are calculated from the electric field given by the PIC simulation through the angular spectrum method. The background color scale shows the laser amplitude, and the white arrows represent the electric field vectors.}
\end{figure}

Figures~\ref{fig:num3}(a)-(c) present the distributions of electric field vectors and amplitude when the generated RP beam exits the plasma and propagates in vacuum after $50\rm{\mu m}$, $100\rm{\mu m}$ and $200\rm{\mu m}$, respectively.  These results are calculated from the electric field given by the PIC simulation corresponding to Fig.~\ref{fig:num2}(b) through the angular spectrum method\cite{Goodman2005}. In comparison with Fig.~\ref{fig:num2}(b), it hints that the intensity modulation initially near the line of $z=0$ ($y<0$) gradually decreases. This may be due to the stronger diffraction of the above higher-order-mode imperfections compared with the main component of the RP beam. These results further illustrate the practicability of our scheme in generating CV beams.

The above magnetized plasma can also be utilized for the generation of vortex beam\cite{Shi2014,Vieira2016,Leblanc2017,QuKenan2017,Long2020}. When the incident beam is circularly polarized, according to Eq.~(\ref{eq:num1}), the laser exiting from the ${{B}_{x}}={{B}_{ext}}\theta$ magnetized plasma can be written as
\begin{eqnarray}
{\bm{E}}_{out}=\left( {{\bm{e}}_{\bm{y}}}+{{\sigma }_{x}}i{{\bm{e}}_{\bm{z}}} \right){{E}_{0}}e^{  i\left( {{k}_{0}}L\sqrt{1-\frac{{n}_{e}}{{n}_{c}}} -\omega t -\eta{\sigma }_{x}\theta \right)},
\label{eq:num5}
\end{eqnarray}
where $\eta = {e{{k}_{0}}L{n}_{e}}B_{ext}/({2{{m}_{e}}\omega {n}_{c} \sqrt{1-{{n}_{e}}/{{n}_{c}}}})$. The output beam is still circularly polarized but with a spiral phase $\exp ( -i{\eta{\sigma }_{x}} \theta )$, which indicates that the exiting laser has been converted into a vortex beam with azimuthal index $l=-{\eta{\sigma }_{x}}$\cite{Allen1992}. 

\begin{figure}
\centering\includegraphics[]{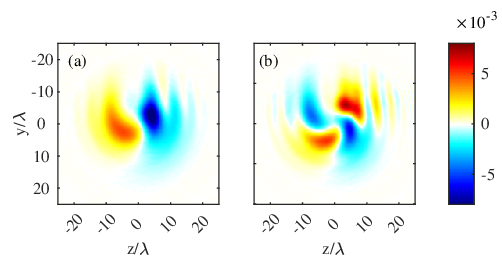}
\caption{\label{fig:num4}Distributions of electric field ${{E}_{z}}$ (normalized by ${{m}_{e}}\omega c/e$) after right-hand circularly polarized (${{\sigma }_{x}}=1$) Gaussian beams pass through different magnetized plasmas with (a) $\eta=1$, (b) $\eta=2$. Other parameters are the same with those in Fig.~\ref{fig:num2}.}
\end{figure}

Figure~\ref{fig:num4}(a) presents the electric field ${{E}_{z}}$ distribution obtained by PIC simulation when a right-hand circularly polarized (${{\sigma }_{x}}=1$) laser exits the above magnetized plasma ($\eta=1$). The two-flap distribution indicates that the existing laser is mainly dominated by the vortex mode with $|l|=1$. By performing the Laguerre-Gaussian (LG) mode decomposition\cite{Shi2014}, the output laser can be predominantly described as $\sqrt{0.04}{{e}^{2.28i}}L{{G}_{0,0}}+\sqrt{0.66}{{e}^{1.41i}}L{{G}_{0,-1}}+\sqrt{0.09}{{e}^{0.77i}}L{{G}_{1,-1}}$. Here, $\sqrt{a_{p,l}}{{e}^{i\phi}}{LG}_{p,l}$ refers to the LG mode with an azimuthal index $l$ and a radial index $p$. It is noteworthy that the waist radius of the LG modes is selected to match that of the incident Gaussian mode. This mode decomposition highlights that an incident Gaussian beam can be effectively converted into a vortex beam with topological charge of $l=-1$. Increasing the parameter $\eta$ allows for the generation of vortex beams with higher values of $|l|$. Figure~\ref{fig:num4}(b) displays the distribution of the electric field ${{E}_{z}}$ when a right-hand circularly polarized (${{\sigma }_{x}}=1$) beam exits the magnetized plasma with $\eta=2$. By making the mode decomposition, it is found that the existing laser is dominated by the LG mode with $l=-2$.


\begin{figure}[b]
\centering\includegraphics[width=8cm]{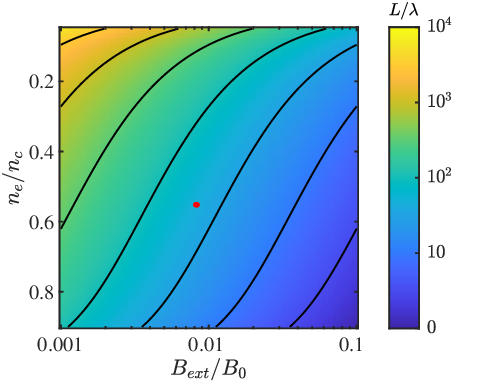}
\caption{\label{fig:num7}Contour plot of the plasma length $L$ (normalized by $\lambda$) with different plasma densities $n_e$ (normalized by $n_c$) and strengths of the axial magnetic field ${B}_{ext}$ (normalized by $B_0 ={m}_{e}\omega /e $).The red point represents the parameters used in the above PIC simulations.}
\end{figure}

In the above PIC simulations, a relatively strong magnetic field ${{B}_{ext}}\approx80\,\rm{T}$ and high plasma density ${{n}_{e}}=6.27\times {{10}^{26}}\,\rm{{m}^{-3}}$ are applied to reduce the simulation cost. Indeed, the requirement of the strength of the magnetic field and plasma density can be significantly reduced. Figure~\ref{fig:num7} shows the dependence of the normalized plasma length $L/\lambda$ on the normalized plasma density $n_e/n_c$ and the strength of the normalized magnetic field $B_{ext}/B_0$. This indicates that a longer plasma length is beneficial for reducing the magnetic field strength as well as the plasma density. For lasers with $\lambda =1\rm{\mu m}$, if the plasma length is chosen as $L=2\,\rm{c m}$ and the plasma density is ${{n}_{e}}=4.46\times {{10}^{25}}\,\rm{{m}^{-3}}$, the required magnetic field is only approximately $4\,\rm{T}$ and is easily achieved experimentally.


In our approach, the generation of an axial magnetic field with ${B}_{x}={B}_{ext}\theta$ is a crucial aspect. We consider a magnetic field that solely possesses an axial component ${{B}_{x}}$, without any azimuthal (${{B}_{\theta }}$) or radial (${{B}_{r}}$) components. It can be readily verified that this  magnetic field satisfies the divergence equation $\nabla \cdot \mathbf{B}=0$. To fulfill the curl equation $\nabla \times \mathbf{B}={{\mu }_{0}}\mathbf{J}$, it is necessary to introduce a current $\mathbf{J}={{B}_{ext}}\sum\limits_{n=1}^{\infty }{n{{a}_{n}}\cos (n\theta )/({{\mu }_{0}}r)}\,{{\mathbf{e}}_{\mathbf{r}}}$, where ${{a}_{n}}=\int_{-\pi }^{\pi }{\theta \sin ( n\theta ){\mathrm{d}}\theta /\pi }=-2{{(-1)}^{n}}/n$. This current can be regarded as the source term responsible for the generation of the ${{B}_{x}}={{B}_{ext}}\theta$ axial magnetic field.

\begin{figure}[t]
\centering\includegraphics[width=8.5cm]{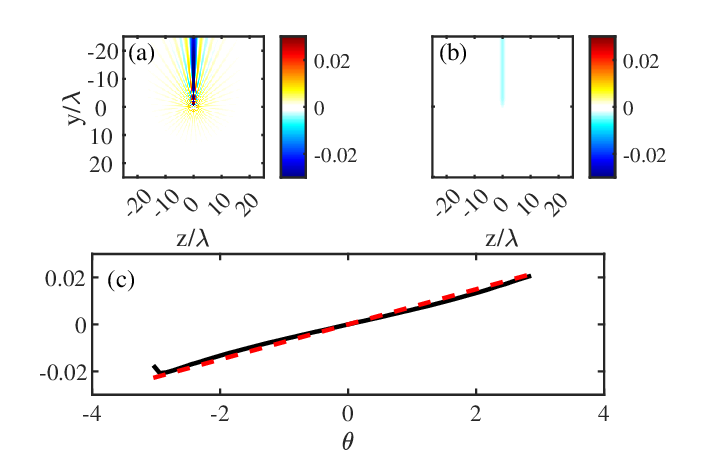}
\caption{\label{fig:num5}(a) Distribution of current ${{J}_{r}}/{J_0}$ obtained by theoretical calculation; (b) Distribution of a semi-infinite plane current ${{J}_{r}}/{J_0}$ ($\approx -{J_y}/{J_0}$); (c) Comparison of magnetic fields (normalized by ${{m}_{e}}\omega /e$) generated by the semi-infinite plane current (black solid line) and theoretical magnetic field (red dashed line) at $r={w}_{0}=10\lambda $.}
\end{figure}

Figure~\ref{fig:num5}(a) displays the distribution of this current (in radial direction) when $n$ is taken until 40. It indicates that this current is mainly distributed near the line of $z=0$ ($y<0$). However, it is worth noting that the current tends to approach infinity as $r$ approaches 0, making it practically impossible to obtain experimentally. To overcome this limitation, we propose employing a semi-infinite flat plate current that addresses the characteristics of the current. This current can be represented as
\begin{eqnarray}
{J_y}/{J_0}=\frac{\pi\lambda{B_{ext}}}{{{B_0}{L_z}}{\cosh}^2(z/{L_z})[1+\exp(y/\lambda)]},
\label{eq:num7}
\end{eqnarray}
where $J_0={{n}_{c}}ec$, $B_0={{m}_{e}}\omega /e$, $L_z$ denotes the characteristic thickness of this current in $z$ direction. The distribution of this current when $L_z=\lambda$ is presented in Fig.~\ref{fig:num5}(b). Figure~\ref{fig:num5}(c) compares the magnetic field generated by this planar current at $r=10\lambda$ with the magnetic field ${{B}_{x}}={{B}_{ext}}\theta$. It is evident that the two magnetic field lines overlap significantly. We employed this magnetic field into the PIC simulation to examine the transformation of a circularly polarized Gaussian beam into a vortex beam. By performing the LG decomposition on the output beam, we obtained $\sqrt{0.07}{{e}^{1.81i}}{{LG}_{0,0}}+\sqrt{0.63}{{e}^{1.34i}}{{LG}_{0,-1}}+\sqrt{0.08}{{e}^{0.7i}}{{LG}_{1,-1}}$, which are nearly identical to those obtained previously.

\begin{figure}[t]
\centering\includegraphics[width=8.5cm]{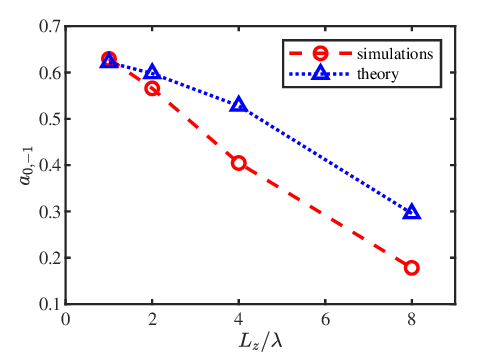}
\caption{\label{fig:num6}The variation of the ${LG}_{0,-1}$ mode energy occupancy ratio (${a_{0,-1}}$) of the outgoing beam with respect to the current sheet thickness ($L_z$). The data represented by red circles is obtained from PIC simulations, while by blue upward-pointing triangles is provided by a simplified theory.}
\end{figure}

It is worth noting the similarity between the current described in Eq.~(\ref{eq:num7}) and the Harris current in the magnetic reconnection process\cite{Harris1962,Yamada2010}, especially in cases involving with the spatially confined X-line in the current direction\cite{Liu2019,Huang2020}. This resemblance implies that the passage of an electromagnetic wave through a magnetic reconnection region can potentially result in the generation of POAM and carry important information along its path. Considering the magnetic reconnection in solar flares, typical parameters are magnetic field strength ${{B}_{ext}}\sim0.01\,\rm{T}$, plasma density ${{n}_{e}}\sim3\times {{10}^{15}}\,\rm{{m}^{-3}}$ and plasma length $L\sim10^8\rm{m}$\cite{Cassak2017}. According to Eq.~(\ref{eq:num4}), for electromagnetic waves with wavelengths greater than $20 \,\rm{\mu m}$, it is possible to be detected as carrying POAM. This result can be confirmed in principle. In addition, the magnetic reconnection in solar flares can now be studied in the laboratory\cite{Zhong2010} based on the scaling laws\cite{Ryutov2000}. Commonly used parameters in the investigation of magnetic field reconnection based on laser plasma processes are ${{B}_{ext}}\sim100\,\rm{T}$ and ${{n}_{e}}\sim5\times{{10}^{25}}\,\rm{{m}^{-3}}$. For laser with $\lambda =800\,\rm{n m}$, the plasma length required to produce significant POAM is about $1\,\rm{mm}$, which is experimentally achievable.

The POAM carried by the outgoing electromagnetic waves holds potential for diagnosing and observing relevant physical quantities. In the case of magnetic reconnection with a
spatially confined X-line extent\cite{Liu2019,Huang2020}, the magnetic fields at the two ends of the current sheet undergo opposite azimuthal changes, leading to the generation of opposing POAMs. This property can be employed for the diagnosis of the length of X-line. Furthermore, the thickness of the current sheet is a significant aspect in the investigation of magnetic reconnection. We conducted a simple study on the conversion of circularly polarized beam passing through magnetized plasma with varying current sheet thicknesses $L_z$. Figure~\ref{fig:num6} displays the dependence of ${LG}_{0,-1}$ mode energy occupancy ${a_{0,-1}}$ on $L_z$, as determined through PIC simulations and a simplified theory. In the PIC simulations, the parameters remain consistent with the previous setup, except for adjusting the current sheet thickness $L_z$. The simplified theory results are obtained by directly incorporating the phase induced by the magnetized plasma into the incident Gaussian beam, followed by the LG mode decomposition. It is evident that the percentage of ${LG}_{0,-1}$ decreases as $L_z$ increases. These findings suggest the potential use of POAM in diagnosing\cite{Torner2005,Berkhout2008} the thickness of the current sheet. However, it must be noted that the results are also influenced by the plasma length and the beam waist of the incident beam, which require further investigation.

Besides the magnetic reconnection process, there are other physical processes or objects in the universe that have magnetic fields near them. For example, the magnetic field near a magnetar can be as strong as $10^8\,\rm{T}$, and whether POAM can be obtained when electromagnetic waves pass through this region is worth further investigation.


In conclusion, a scheme converting the incident intense linearly polarized Gaussian beam into a cylindrical vector beam is proposed, which is achieved by setting up an axial magnetic field with azimuthal distribution ${{B}_{x}}={{B}_{ext}}\theta $. Three-dimensional particle-in-cell simulation results verify the feasibility and efficiency of this scheme. Besides, when considering circularly polarized Gaussian beam passing through the proposed magnetized plasma, vortex beam with $\left| l \right|=1$ can be generated. By increasing the strength of the magnetic field ${{B}_{ext}}$ (or plasma length $L$, electron number density ${{n}_{e}}$), arbitrarily higher $\left| l \right|$ mode vortex beams can also be generated. This scheme is free of the optical thermal threshold and is applicable for intense lasers with powers as high as $\sim \rm{PW}$ in some laser facilities.

Our investigation also reveals a promising and novel source of photon orbital angular momentum in astrophysics and space physics. Specifically, we have demonstrated that orbital angular momentum can be acquired by an electromagnetic wave passing through a magnetic reconnection region. This photon orbital angular momentum holds the potential for diagnosing the length of the X-line and the thickness of the current sheet in this process. Further research is required to explore and understand these possibilities in greater depth.

\begin{acknowledgments}
This research was supported by the National Natural Science Foundation of China (NSFC) under Grant No. 11975014, by the Strategic Priority Research Program of Chinese Academy of Sciences, Grant Nos. XDA25050400 and XDA25010200. The authors would like to thank and acknowledge Q. Lu and K. Huang for helpful discussions.
\end{acknowledgments}

\nocite{*}

\bibliography{CVBbyIF}

\end{document}